# UNIFICATION AFTER 150 YEARS

## S. Sahoo[1] and M. Goswami[2]

[1]Department of Physics, National Institute of Technology,
Durgapur – 713209, West Bengal, India.
E-mail: sukadevsahoo@yahoo.com

[2]Department of Physics, Regional Institute of Education (NCERT),
Bhubaneswar - 751022, Orissa, India.
E-mail: manasigoswami1@yahoo.com

**Abstract**

In 1861, James Clerk Maxwell unified electricity, magnetism and light. Today physicists are trying to unify everything else theoretically as well as experimentally with a view to develop a fundamental theory.

## 1. Introduction

The primary goal of physics is to understand the wonderful variety of nature in a unified way. Exactly 150 years ago, James Clerk Maxwell published a paper in the *Philosophical Magazine* with a title, 'On Physical Lines of Force'. Between 1861 and 1862 he unified apparently three separate phenomena – electricity, magnetism and light – known as electromagnetism [1–4]. According to Maxwell, magnetic fields can be generated by varying electric fields as well as by travelling charges. For expressing this, he established a set of equations those can apply to electricity and magnetism in disciplines ranging from astronomy to telecommunications [3]. The application of Maxwell's equations in materials depends on how those materials affect the propagation of electric and magnetic fields. These equations can be applied not only to materials but also to the empty space. Maxwell observed a single propagation velocity in empty space for all electromagnetic radiation – known as the *speed of light*. Now it is also applied to radio waves, infrared, ultraviolet waves, X-rays and $\gamma$-rays. Later on Einstein realised that the speed of light in empty space is constant and it is independent of the position and the direction of motion of the observer. Maxwell's unification 'not only explained all known electromagnetic phenomena, it explained light and pointed to the existence of kinds of radiation' [2]. According to Feynmann, 'the most significant event of the nineteenth century will be judged as Maxwell's discovery of the laws of electrodynamics'.

Paul Dirac combined special theory of relativity and quantum theory to create his famous equation known as Dirac equation, which predicted the existence of antimatter [5,6]. Feynman, Schwinger and Tomonoga created the first relativistic quantum field theory, quantum electrodynamics (QED), which explained the interaction of light and matter. The development of quantum field theories emphasised the importance of symmetries and their links to conservation laws and unification. In 1918 the mathematician, Emmy Noether proved that every symmetry of nature yields a conservation law; conversely, every conservation law reveals an underlying symmetry [4]. A theory for quark interactions was constructed by analogy with QED. This theory was called quantum chromodynamics (QCD). This theory postulated the existence of massless particles called gluons by which the quarks are held or glued together. In 1960s the weak interaction and the electromagnetism – these two theories were unified by Abdus Salam, Steven Weinberg and Sheldon Glashow and formed electroweak theory. This theory predicted the existence of heavy force carrying particles along with the massless photon, and these (the $W^\pm$ and $Z^0$ particles) were discovered at CERN in 1983. Carlo Rubbia and Simon Vander Meer received the Nobel Prize in Physics in 1984 for this discovery. In the 1970s a new symmetry, called supersymmetry [7,8] was discovered. Supersymmetry introduces a fermionic (bosonic) partner for every boson (fermion) differing by half a unit of spin quantum number. This partner is called superpartner (sparticle) of the origional particle. By combining the symmetries of electroweak and colour interactions theorists proposed many grand unified theories (GUTs) involving supersymmetry. This led to the prediction of new supersymmetric particles accompanying all the known particles. In 1976, supergravity [9] was proposed and attempts were made to combine it with supersymmetric versions of the standard model to stand as quantum gravity (Figure1) [10].

Electricity + Magnetism + Optics = Electromagnetism
Electromagnetism + Quantum theory = QED
QED + Weak interaction = Electroweak theory
Strong interaction + Quantum theory = QCD
QCD + Electroweak theory = GUT
GUT + Gravity = Quantum gravity (String theory ?)

The two theories, the electroweak theory and the QCD, form the standard model (SM) [11,12] of elementary particles. This model assumes three generations (or families) of quarks and three generations of leptons. Quarks are called (up, down), (charm, strange), and (top, bottom). Leptons are called (electron, electron-neutrino), (muon, muon-neutrino) and (tau, tau-neutrino). All neutrinos are assumed to be massless and neutral. The first generation quarks and leptons are found in ordinary matter. But second and third generation quarks and leptons are found only in cosmic rays and accelerators. The final particle of the SM, the Higgs boson (H), has not been discovered yet. It is very important because it is responsible for the mechanism (Higgs mechanism [13,14]) by which all other particles acquire mass. The Higgs discovery could take the unification a giant step by filling in the last and most crucial piece of the SM. It is expected that the Higgs boson will be produced and studied at the Large Hadron Collider (LHC), the world's largest particle accelerator.

The SM unifies the strong, electromagnetic and weak forces. It does not take into account the gravitational force. That is why physicists are not satisfied with the unification at this stage. The ultimate goal is to construct a unified theory that would reveal how all observed particles and forces are just different manifestations of a single underlying system, which can be expressed within a common mathematical framework [1]. The gravitational force is based on the Einstein's general theory of relativity. This theory is based on the principles of classical mechanics and not of quantum mechanics. Since other forces in nature obeys the rules of quantum mechanics, any theory that attempts to explain gravity as well as the other forces of nature must satisfy both gravity and quantum mechanics. Quantum mechanics works at very small distances. For example, the physics of molecules, atoms, nuclei and elementary particles can be described within the framework of quantum mechanics. But general theory of relativity describes gravity at large distances. For example, the physics of the solar system, neutron stars, black holes and the universe itself can be described within the framework of general theory of relativity. It is seen that string theory [15–17] works very well at large distances where gravity becomes important as well as at small distances where quantum mechanics is important. This is why string theory comes into picture.

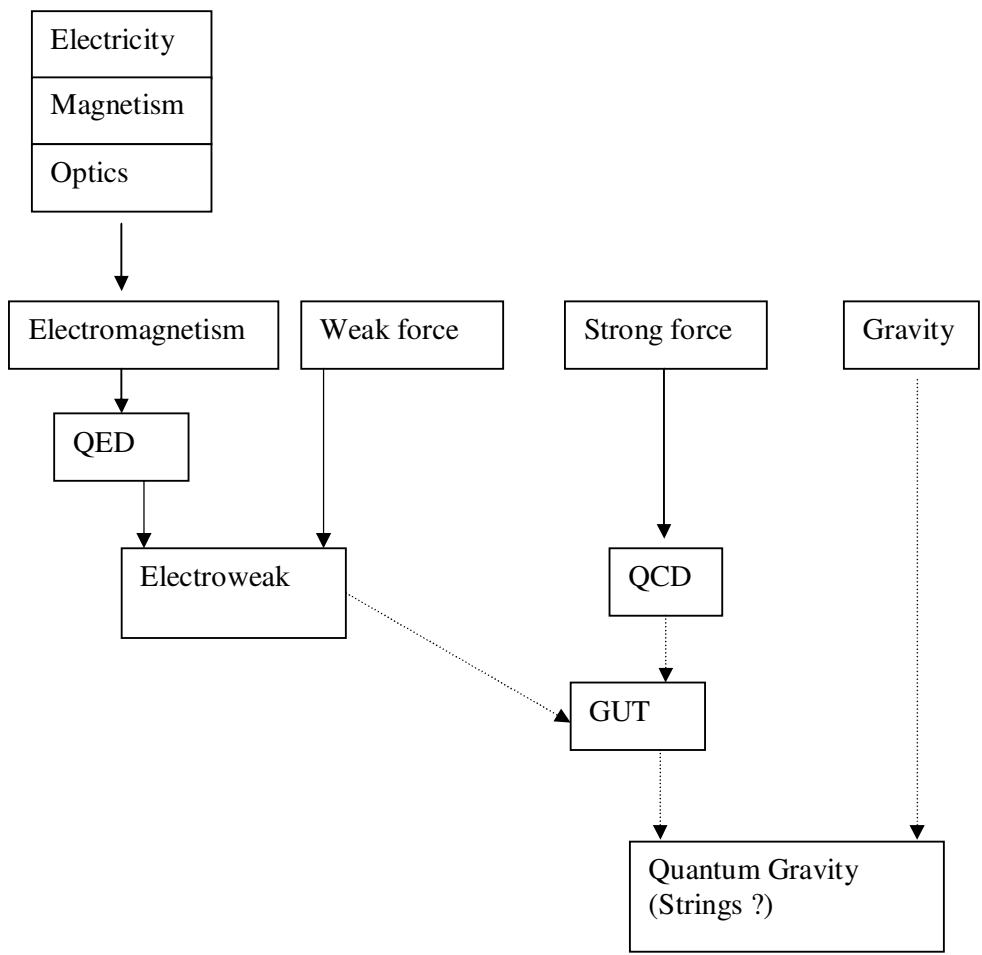

**Figure 1: Unification of fundamental forces**

String theory is an attempt to unify all the four fundamental forces of nature. In the standard model the elementary particles are mathematical points. But in string theory, instead of many types of elementary point-like-particles, we assume that in nature there is a single variety of one-dimensional fundamental objects known as strings [18]. It is not made up of anything but other things are made up of it. Like musical strings, this basic string can vibrate, and each vibrational mode can be viewed as a point-like elementary particle, just as the modes of a musical string are perceived as distinct notes. String theory [19,20] provides a framework to address some fundamental issues in cosmology and elementary particle physics. Examples are dark matter, dark energy, supersymmetric particles, unification of all interactions, etc. In 2005, the scientists in the Relativistic Heavy Ion Collider (RHIC) at Brookhaven National Laboratory in New York discovered that string theory could be as useful as QCD in explaining the strong nuclear forces involved in a quark-gluon plasma. Recently, it is also claimed that string theory can be applied in condensed matter physics. It is claimed that string theory will explain the origin of high-temperature superconductivity [21]. A type of black hole predicted by string theory may help to explain the properties of a mysterious class of materials called 'strange metals' [22]. String theory can also be applied to describe the behaviour of hot particle plasma and supercooled lattices of atoms. For a long time string theorists have proposed an idea that there should be only one fundamental theory which describes our world and that is our string theory. The completeness of unification in physics requires the real existence of Higgs boson, supersymmetry, and extra dimensions etc which are discussed in the next sections very briefly.

**2. Higgs boson**

The Higgs boson is a hypothesized particle, which if it exists, would give the mechanism by which particles acquire mass. The standard model suggests that just after the big bang all particles were massless. As time passed on, the universe cooled and temperature fell below a critical value, an invisible field called the 'Higgs field' filled all space [23]. The particle associated with the Higgs field is called the Higgs boson. Since the Higgs field is a scalar field, the Higgs boson has no spin, and hence no intrinsic angular momentum. The Higgs boson is also its own antiparticle and is CP-even. One of the important properties of this field is that the Higgs field is exactly the same everywhere whereas the magnetic or gravitational fields vary from place to place. When particles are moving in a uniform Higgs field, they change their velocities i.e. they accelerate. The Higgs field exerts a certain amount of resistance or drag, this is the origin of inertial mass.

The Higgs boson interacts with or couples to elementary particles proportionally to their masses [24]: the more massive is the particle, the stronger is its interaction with the Higgs boson. The Higgs boson has not been discovered yet. The value of the Higgs mass is not specified in the SM. From both the direct and indirect measurements the entire mass range of the Higgs boson is expected as 114.4 GeV $\leq m_H \leq$ 1 TeV. The LHC at CERN is expected to measure the mass of Higgs boson accurately. But there is no guarantee that the LHC will find it [23]. Recently, in late April, 2011 results from the LHC seemed to show a preliminary signal of the Higgs boson [25]. John Ellis, a theoretical physicist at King's College London says, it is entirely possible that the LHC will turn up not one Higgs particle, but a whole family of them. The Higgs boson gives mass to every elementary particle which has mass, including the Higgs boson itself. That is why it is often called as "the God particle" [26], after the title of Leon Lederman's book, *The God Particle: If the Universe is the Answer, What is the Question?*.

**3. Supersymmetry**

Supersymmetry is a hypothetical symmetry between fermions and bosons. Unlike traditional symmetries, supersymmetry does not treat bosons and fermions as two different classes of particles. The supersymmetry operation converts bosons into fermions and vice versa. For each particle, it predicts the existence of a superpartner (hence, doubles the SM particle spectrum) which should have the same properties but with a spin different by a unit ½ and also a different mass as supersymmetry must be broken in nature. As a matter of convention, fermionic superpartner of known bosons are written by the suffix *-ino* (for example, "gravitino, photino, gluino" etc.); the bosonic superpartners of fermions are written by a pre-fixed *s-* (for example, "squark, slepton" etc.). The discovery of any such bosinos and sfermions would confirm the important prediction of superpartners, which is common to all supersymmetric models. For every particle, supersymmetry predicts a partner (sparticle) with spin differing by one half [27]. Some particles, and their superpartners [10,28] are given in Table 1.

**Table 1**

| Matter particle | Spin | Superpartner | Spin |
|---|---|---|---|
| Electron | ½ | Selectron | 0 |
| Muon | ½ | Smuon | 0 |
| Tau | ½ | Stau | 0 |
| Neutrino | ½ | Sneutrino | 0 |
| Quark | ½ | Squark | 0 |
| **Force-carrier** | **Spin** | **Superpartner** | **Spin** |
| Graviton | 2 | Gravitino | 3/2 |
| Photon | 1 | Photino | ½ |
| Gluon | 1 | Gluino | ½ |
| $W^\pm$ | 1 | *Wino*$^\pm$ | ½ |
| $Z^0$ | 1 | Zino | ½ |
| Higgs | 0 | Higgsino | ½ |

Supersymmetry helps in the grand unification of the strong, weak and electromagnetic forces, leaving gravity alone. In the SM, the strong force's interaction strength is different from the strengths of the weak and electromagnetic forces. But if supersymmetry is assumed to really exist, quantum corrections show that all the three strengths to be exactly equal, which is expected if these forces are actually one or unified. Supersymmetry may provide an explanation for dark matter problem in cosmology and provides some dark matter candidates like neutralino, gravitino etc. Supersymmetry protects the Higgs mass from acquiring large values [24]. The theoretical proposition of supersymmetry needs to be experimentally tested i.e. sparticles of elementary particles must be observed in high-energy collider experiments. It is hoped that supersymmetry will experimentally be tested at colliders: Large Hadron Collider (LHC) and International Linear Collider (ILC) [29]. Although it is a debatable question whether nature possesses supersymmetry or not, there are many fruitful applications of supersymmetry both in the area of high-energy particle physics and in quantum mechanics.

**4. Extra dimensions**

String theory can be formulated in 10-dimensional space-time rather than the 4-dimensional space-time (3 spatial and 1 time dimension) in which we live. So physicists require extra dimensions for unification beyond the standard model. According to Kaluza and Klein, if we assume that 6 spatial dimensions are curled up into tiny sizes, while the remaining 3 dimensions extend to infinity (or at least to very large distances) then 9-dimensional space will be seen as 3-dimensional space. This process is known as compactification [15,30]. But unfortunately such extra dimensions have not been observed directly till today. These hidden dimensions are still in hiding [31]. The LHC could detect those extra dimensions if the particles generated by collisions have enough energy and therefore short enough wavelengths to start spiralling around those tightly curled dimensions. Albert de Roeck, the deputy spokeman of the Compact Muon Solenoid (CMS) experiment at the LHC, says that if the size of the extra dimension is smaller than $10^{-19}$ metres, then the energy required to probe them will be beyond the LHC's reach [1]. Again Michael Duff, a physicist at Imperial College London has said that if there are extra dimensions they are at the Planck scale of $10^{-35}$ metres. So there is no guarantee that the LHC will find them. Still then we hope for positive result.

**5. Conclusions**

Search for a unified theory is not completed till today. Even if Einstein, the greatest physicists of the 20$^{th}$ century, spent his last thirty years (which is more than half of his scientific career) in search of a unified theory of physics [32]. But he was not fully successful. He himself admitted in 1954, "*I must seem like an ostrich who buries its head in the relativistic sand in order not to face the evil quanta*". Now we hope that at very high energies, gravitational force will unite with the other three fundamental forces into a single 'superforce'. If this hypothesis is correct then during the first instants of the big bang the universe was dominated by this unified superforce. Thus, the microscopic world of high-energy particles and fundamental forces is extremely linked to the large scale structures of the universe. Many physicists think that the "string theory" is the most promising candidate theory for a unified description of the fundamental particles and forces existing in the nature including gravity.

String theory is an attempt to unify all the four fundamental forces by modelling all particles as various vibrational modes of an extremely small entity known as *string*. All the elements are there: quantum mechanics, bosons, fermions, gauge groups and the gravity, and in this sense be "quantum gravity". This theory requires six extra spatial dimensions beyond the real 4-dimensional world in which we live. It is assumed that these extra dimensions are curled up into so small sizes that we are unable to detect its existence directly. Till today there is no experimental confirmation of string theory [20]. Still now there are two big problems [33,34] in the string theory: (i) Typical size of the string is of the order $10^{-33}$ cm. In order to see this structure we need extremely powerful microscopes, or equivalently, extremely powerful accelerators which can accelerate particles to an energy of order $10^{19} m_p c^2$, where $m_p$ is the mass of a proton at rest. But at present we have succeeded in accelerating particles up to only an energy of about $10^3 m_p c^2$. This is why the size of strings remains beyond the dreams of experimental physics for the near future i.e. big problem for small size [33]. (ii) At present, we do not even know the full equations of string theory and its complexity leads to a large number of possible solutions. But we do not know the appropriate criteria for selecting those that apply to our universe.

The fate of the string theory will be determined by its experimental predictions. We hope the LHC / ILC will search for the existence of string theory. If the string theory is verified in experiments, it will completely revolutionize our concepts of space, time and matter. It will reveal secrets of the universe [35]. The search for a unified theory is indeed a valuable one, and should be continued in order to further the understanding of all humankind. Although 150 years have already passed after Maxwell's unification, the unification process is not completed yet. It is expected that the work of unification [36] will be completed by 2050. But can we actually do it? This will not be the end of physics. But it will mark the end of a certain kind of physics: the search for a unified theory that entails all other facts of physical science.

*References*

1. M. M. Waldrop, *Nature,* **471**, 286 (2011).
2. P. Balaram, *Curr. Sci.*, **100**(8), 1117 (2011).
3. Editorials, *Nature,* **471**, 265 (2011).
4. A. Beiser, *Concepts of Modern Physics*, 6[th] Edition, Tata McGraw-Hill Company Limited, New Delhi (2004).
5. P. A. M. Dirac, *Proc. R. Soc.*, **A117**, 610 (1928); P. A. M. Dirac, *Proc. R. Soc.,* **A118**, 351 (1928).
6. S. Sahoo, R. K. Agrawalla and M. Goswami, *Physics Education (IAPT)*, **23**(4), 273 (2007).
7. S. Sahoo and M. Goswami, *IAPT Bulletin*, **24**(5), 140 (2007); M. E. Peskin, arXiv: 0801.1928 (2008).
8. S. K. Soni and Simmi Singh, *Supersymmetry*, Narosa Publishing House, India (2000); G. Kane and M. Shifman, *The Supersymmetric World*, World Scientific Publishing Co. Pte., Singapore (2000).
9. M. Jacob, *Supersymmetry and supergravity*, Netherlands and World Scientific Publishing Co. Pte., Singapore (2000).

10. S. Adams, *Frontiers Twentieth-Century Physics*, Talor and Francis Ltd, London (2000); D. Griffiths, *Introduction to Elementary Particles*, John Wiley and Sons, Singapore (1987).
11. R. M. Godbole and S. Mukhi, *Resonance*, **10**(2), 33 (2005); A. Sen, *Resonance*, **10**(12), 86 (2005).
12. S. Sahoo, *IAPT Bulletin*, **25**(7), 228 (2008); S. Sahoo, *Physics Education (IAPT)*, **22**(2), 85 (2005); S. Sahoo and M. Goswami, *Physics Education (IAPT)*, **23**(2), 113 (2006).
13. P. W. Higgs, *Phys. Lett*. **12**, 132 (1964).
14. P. W. Higgs, *Phys. Rev*. **145**, 1156 (1966).
15. A. Sen, *Curr. Sci.*, **89**(12), 2045 (2005); A. Sen, *Curr. Sci.*, **81**(12), 1561 (2001); A. Sen, *Curr. Sci.*, **77**(12) 1635 (1999).
16. B. Zwiebach, *A First Course in String Theory,* Cambridge University Press, New York (2004); P. C. W. Davies and J. Brown, *Superstrings*, Cambridge University Press, London (1997).
17. J. Polchinski, *String Theory Vol. I – An Introduction to the Bosonic String Theory,* Cambridge University Press, UK (2004); J. Polchinski, *String Theory Vol. II – Superstring Theory and Beyond*, Cambridge University Press, UK (2004).
18. J. H. Schwarz, *Curr. Sci.*, **81**(12), 1547 (2001).
19. Spenta R. Wadia, *Curr. Sci.*, **95**(9), 1252 (2008).
20. M. Dine, *Physics Today*, **60**(12), 33 (2007); M. Dine, *Supersymmetry and String Theory: Beyond the Standard Model*, Cambridge University Press, New York (2007).
21. E. Hand, *Nature*, doi:10.1038/news.2009.699, published online 19 July 2009.
22. E. S. Reich, *Nature*, doi:10.1038/news.2010.547, published online 19 October 2010.
23. M. Cid and R. Cid, *Physics Education*, **45**(1), 73 (2010) (IOP Publishing).
24. A. Djouadi, in the book *Mass and Motion in General Relativity* edited by L. Blanchet, A. Spallicci and B. Whiting, Springer, pp. 1 – 23 (2011).
25. G. Brumfiel, *Nature*, **473**, 136 (2011).
26. G. M. Helaluddin, *IAPT Bulletin*, **2**(4), 96 (2010).
27. E. Gallo, *Int. J. Mod. Phys. A,* **22**, 5513 (2007) [arXiv: hep-ex/0612028].
28. M. Chaichian and R. Hagedorn, *Symmetries in Quantum Mechanics*, Institute of Physics Publishing, London (1998); K. Cahill, hep-ph/9907295 (1999); J. Beckers, N. Debergh and A. G. Nikitin, *J. Math. Phys.* **33**, 152 1992) [arXiv: math-ph/0508021].
29. J. Kalinowski, hep-ph/0508167 (2005).
30. M. Atiyah, *Curr. Sci*. **89**(12), 2041 (2005).
31. S. Sahoo, *Physics Teacher (IPS)*, **46**(1-2), 28 (2004).
32. D. Gross, *Curr. Sci.*, **89**(12), 2035 (2005).
33. S. Sahoo, *Eur. J. Phys.* **30,** 901 (2009).
34. S. Sahoo and R. K. Agrawalla, *IAPT Bulletin*, **25**(9), 293 (2008).
35. Y. S. Cho and S. – T. Hong, arXiv:1010.3485 [physics.gen-ph] (2010).
36. S. Weinberg, *Scientific American*, pp. 4 – 11 (Special edition March 2003).